\documentclass[aps,twocolumn,showpacs,preprintnumbers,amsmath,amssymb,nofootinbib,superscriptaddress,showkeys,prd]{revtex4-1}

\usepackage{epsfig}
\usepackage{graphicx}
\usepackage{mathptmx}      % use Times fonts if available on your TeX system
\usepackage{latexsym}

\def\slashchar#1{\setbox0=\hbox{$#1$}
   \dimen0=\wd0 \setbox1=\hbox{/} \dimen1=\wd1
   \ifdim\dimen0>\dimen1 \rlap{\hbox to \dimen0{\hfil/\hfil}} #1
   \else  \rlap{\hbox to \dimen1{\hfil$#1$\hfil}} / \fi}

\begin{document}

\title{Regge trajectories of Excited Baryons, quark-diquark
  models and quark-hadron duality}

\author{Pere Masjuan} \email{masjuan@ifae.es}
\affiliation{Grup de F\'{\i}sica Te\`orica, Departament de
  F\'{\i}sica, Universitat Aut\`onoma de Barcelona  and Institut de
  F\'{\i}sica d'Altes Energies (IFAE), The Barcelona Institute of
  Science and Technology (BIST), Campus UAB,\\ E-08193 Bellaterra
  (Barcelona), Spain}

\author{Enrique Ruiz Arriola}\email{earriola@ugr.es}
\affiliation{Departamento de
  F\'{\i}sica At\'omica, Molecular y Nuclear
and Instituto Carlos I de F{\'\i}sica Te\'orica y Computacional \\
Universidad de Granada, E-18071 Granada, Spain.}

\date{\today}

\begin{abstract}
The parton model relations in conjunction with quark-hadron duality in
deep inelastic scattering suggests an asymptotic dominance of
quark-diquark type of Baryonic excited states with a radial Regge
uniformly distributed mass squared spectrum,  $M_{n}^2 = \mu^2 n +
M_0^2$. We argue that this points to a linearly quark-diquark confining
potential. We analyze the radial ($n$) and angular-momentum ($J$)
Regge trajectories for all light-quark states with baryon number one
listed in the 2016 edition of the Particle Data Tables.  The
parameters of the mass squared trajectories are obtained by linear
regression assuming $\Delta M_n^2 \sim M_n \Gamma_n $ weighted with
the width $\Gamma_n$ of the resonance and the error analysis is
carried out accordingly.
\end{abstract}



\maketitle

\section{Introduction}

One of the unresolved issues in particle physics is the completeness
of the hadronic spectrum. Assuming that Quantum Chromodynamics (QCD)
is the fundamental theory, the main problem is then to establish what
states should be labeled or recorded as ``eigenstates'' of the QCD
Hamiltonian.  While this question may unambiguously be answered in a
finite box thanks to the discretization imposed by the boundary
conditions on the quark and gluon fields, difficulties arise in the
continuum, where scattering experiments and spectroscopy measurements
are actually conducted; in the infinite volume limit the concept of an
eigenstate becomes definitely more fuzzy.

The Particle Data Group (PDG) booklet~\cite{Olive:2016xmw} compiles
the agreed upon result of the many existing phenomenological analyses
and identification of states, mostly resonances, determined from
hadronic reactions and, till now, the classification and listings
faithfully feature the quark model classification of states. This
scheme records and rates single states one-by-one with growing number
of stars, $*,**,***,****$, depending on the increasing estimated
confidence level on their existence~\cite{Crede:2013sze}. The rating
of states changes with time depending on the detailed features of the
analysis~\cite{Klempt:2017lwq}.

One should remind that within a Hamiltonian viewpoint, resonances in
the continuum are identified as the so-called Gamow states which are
not normalizable in the usual Hilbert space sense, as they are not
conventional irreducible representations of the Poincar\`e's
group~\cite{Bohm:2004zi}. Even in the simplest potential scattering
situation the completeness relation involves bound states and the
continuum, which can be rewritten as a discrete sum of the Gamow
states and a remainder which is generally
non-vanishing~\cite{Berggren:1968zz}. This circumstance adds more
difficulties to a practical definition of
completeness~\footnote{Gamow states belong instead to the rigged
  Hilbert space where completeness has a well defined meaning, see
  e.g. Ref.~\cite{delaMadrid:2002cz} for a pedagogical exposition and
  references therein. For the present paper we stay within the
  conventional Hilbert space interpretation.}.

The completeness issue becomes particularly severe in the case of
baryons; the exceedingly many baryonic excited states predicted by the
quark model (where baryons are $qqq$ states)~\cite{Capstick:1986bm}
have apparently not yet been found. This persistent puzzle is refereed
to as the {\it missing resonance problem}~\cite{Capstick:1992uc} (see
also \cite{Hey:1982aj} and \cite{Capstick:2000qj} for reviews and
references therein).  The two commonly accepted, not necessarily
incompatible, possible explanations to the puzzle assume i) Weak
coupling of the predicted states to the particular production process
(photo-production, $\pi N$ scattering etc.) or ii) Dynamical reduction
of degrees of freedom due to diquark clustering (for a review see
\cite{Anselmino:1992vg}).  Both possibilities have triggered a great
deal of experimental as well as theoretical activity, but again based
on {\it individual} one-to-one mapping of resonance states which have
a mass spectrum and which are produced with different
backgrounds. More recently, a revision of the resonance ratings
problem suggests that the missing resonances might indeed be found in
the intricacies of partial wave analyses to a large
database~\cite{Klempt:2017lwq}. Under these circumstances independent
sources of information are most welcome.

In the opposite extreme to the individual resonance approach, a
thermodynamic approach to the completeness problem is more global as
it concerns all states as a whole and can be verified from the study
of the equation of state; at not too high temperatures most resonances
behave as narrow and they can effectively be regarded as particles in
the partition function~\cite{Dashen:1969ep,Dashen:1974jw}.  Some
separation of quantum numbers can be imposed with the study of
susceptibilities of conserved charges, where a combination of
degeneracy and level density is involved (see e.g.
Ref.~\cite{RuizArriola:2016qpb} and references included).

In the present work we return to an intermediate possibility to
address completeness for baryons inspired by the old notion of
Quark-Hadron duality, namely the coupling of QCD quark bilinear
currents to hadrons close to the Bjorken limit. Quark-hadron duality
was first established for inclusive electro-production on the proton
in the deep inelastic scattering (DIS)
regime~\cite{Bloom:1970xb,Bloom:1971ye}.  It was soon found that
scaling in the Bjorken limit implies a mass
formula~\cite{Domokos:1971ds} for the excited baryon spectrum and has
been a recurrent feature in the data ever since (for a review see
e.g. \cite{Melnitchouk:2005zr} and references therein). Such a
framework has the virtue that the parton model requires infinitely
many states with the same quantum numbers, providing a sort of closure
(completeness) relation for narrow resonant states.

In the meson case, where currents connect the vacuum with $\bar q q$
states, the situation is much simpler and quark-hadron duality implies
there a radial Regge trajectory for excited mesons. Building on
previous works~\cite{Anisovich:2000kxa} a complete analysis with
uncertainties incorporating the finite resonance width was carried out
previously for $\bar q q$ states~\cite{Masjuan:2012gc}. In the present
paper we follow a similar strategy to analyze excited baryons.

The paper is organized as follows.  In Section~\ref{sec:scaling} we
review the argument leading to the mass formula inferred from DIS. By
using a semi-classical approximation we identify the mass formula as
the fingerprint of quark-diquark structure of excited
baryons in Section~\ref{sec:diquarks}. The phenomenological analysis of the
radial Regge trajectories is carried out with the upgraded PDG 2016
information in Section~\ref{sec:regge} on the light of their resonant
nature. We also compare to some recent studies in
Section~\ref{sec:comp}. Finally, in Section~\ref{sec:concl} we draw
our main conclusions.

\section{Bjorken Scaling and Baryon Spectrum}
\label{sec:scaling}

We start by reviewing here in a simplified way the main
argument~\cite{Domokos:1971ds} leading to the asymptotic uniformly
distributed mass squared spectrum (spinor details are lengthy and
handled in that work). For unpolarized nucleon targets the forward
Compton scattering in the gauge invariant
form reads~\footnote{ We use the convention for the metric $g_{00}= -
  g_{ii}=1 $, single particle normalization $\langle p' , \lambda'
  | p , \lambda \rangle = (M/E) (2\pi)^3 \delta( \vec p' - \vec p)
  \delta_{\lambda',\lambda}$ with $E= \sqrt{\vec p^2+M^2}$ and Dirac
  spinors as $\bar u u=1$.}
\begin{eqnarray}
W_{\mu \nu} (p,q;s) &=& 
\frac1{4\pi}\sum_\lambda\int d^4x e^{iq\cdot x}\,
\Big\langle p, \lambda \Big| [ J_{\mu}(x),J_{\nu} (0)]
\Big| p, \lambda \Big\rangle\,  \\ 
& = & \left(-g_{\mu \nu} + \frac{q_{\mu} q_{\nu}}{q^2}\right)
W_{1} (\nu,Q^2) \nonumber \\ 
&+&\left(p_{\mu} - q_{\mu}\frac{p\cdot q}{q^2}\right)
\left(p_{\nu} - q_{\nu}\frac{p\cdot q}{q^2}\right)
\frac{1}{M^2} W_{2} (\nu,Q^2) \nonumber 
\end{eqnarray}
with $J_\mu= \bar q \hat Q \gamma_\mu q $ the e.m. current and $\hat
Q=(2/3,-1/3,-1/3)$ the charge operator for (u,d,s) quarks, $p^2=M^2$
with $M$ the nucleon mass, $\lambda$ the nucleon helicity, $\nu = p
\cdot q /M$ and $Q^2=-q^2$.  For the parton model in the Bjorken
limit, $ Q^2\to\infty $ with $x=Q^2/(2p\cdot q) $ fixed, one obtains
both scaling
\begin{eqnarray}
W_1(x,Q^2)\, \to  F_1(x) \, , 
\qquad  
\frac{p\cdot q}{M^2}\,W_2(x,Q^2)\, \to  F_2(x)\, , 
\end{eqnarray}
and the Callan-Gross relation $ F_2(x)= 2 x F_1(x)$ featuring the Spin
1/2 nature of partons, so that we may focus on just one, say $F_1(x)$.
Defining $J_T \equiv J (0)\cdot \epsilon_T $ with $\epsilon_T $ a
four-vector such that $\epsilon_T \cdot \epsilon_T=-1$ and $
\epsilon_T \cdot q = \epsilon_T \cdot p =0$ we get $W_1=\epsilon_T^\mu
\epsilon_T^\nu W_{\mu\nu}$. Inserting a complete set of eigenstates
$|\alpha \rangle$, we get
\begin{eqnarray}
W_1 &=& 
(2\pi)^3 \frac12 \sum_{\alpha,\lambda} |\langle p,\lambda | J_T | \alpha \, \rangle |^2
\delta^4 ( p+q - P_\alpha) \, . 
\end{eqnarray}
Taking $|\alpha \rangle \equiv |p+q,J m \nu n \rangle$ where
$J,m,\nu,n $ specify the angular momentum, magnetic,
normality and {\it radial} quantum numbers respectively for the
resonance $R$ with mass $M_R=M_{J \nu n}$, in our normalization   we have 
\begin{eqnarray}
\sum_\alpha &\to& \sum_{Jm\nu n} \int \frac{d^3 P_R}{(2\pi)^3}
\frac{M_R}{E_R}
\nonumber \\
&=& \sum_{Jm\nu n} \int \frac{d^4
  P_R}{(2\pi)^3} 2 M_R \delta(P_R^2 -M_R^2) \, . 
\label{eq:wmunu-res}
\end{eqnarray}
%hence 
%\begin{eqnarray}
%W_1 &=&\frac12 \sum_{R,\lambda,\lambda'} | \langle N,\lambda | J_T | R,
%\lambda' \rangle|^2 \delta( (p+q)^2 - M_R^2 )
%\end{eqnarray}
In order to see how scaling arises in the
Bjorken limit at the hadronic level consider the invariant
$N+\gamma^*$ squared mass 
\begin{eqnarray}
s = (p+q)^2 = M^2 + Q^2 \left( 1/x-1\right) \, , 
\end{eqnarray}
and {\it assume} that $N \to R$ transition form factors fufills a
scaling relation in terms of the resonance mass, $M_R$, of the form 
\begin{eqnarray}
\sum_{\lambda m}| \langle p,\lambda | J_T |
p+q,J m\nu n\rangle |^2 = \frac{M_R M^2}{Q^2} \left[ G_{J\nu} \left(-\frac{Q^2}{M_R^2} \right) \right]^2 ,  
\end{eqnarray}
We are left with $\sum_{J \nu n}$ and replacing the sum in $n$ by an
integral we can evaluate the Dirac delta function and get a factor
$dn/dM^2 \equiv 1/ (d M^2_{J\nu n} /dn) $,
\begin{eqnarray}
W_1 &=& \sum_{J\nu n} \left[ G_{J\nu} (-Q^2/M_{J \nu n }^2)\right]^2 
\frac{M_{J\nu n}^2}{Q^2}\delta( s - M_{J\nu n}^2) \nonumber  \\
& \to & F_1(x) =   
\frac{1-x}{x} \sum_{J\nu} \left[G_{J\nu} \left(\frac{x}{x-1} \right)\right]^2 \frac{M^2}{d M^2_{J\nu n} /dn} \, . \label{eq:sumint}
\end{eqnarray}
Scaling follows from a constant mass squared level density
\begin{eqnarray}
 d M_{J\nu n}^2/ dn = \mu_{J\nu}^2 \, .
\label{eq:qhd}
\end{eqnarray}
In Ref.~\cite{Domokos:1971ds} it was found that this also implies the
Drell-Yan-West asymptotic relation between the form factor $ G(-Q^2) \to Q^{-n} $ at large $Q$ and the structure function $F_1
(x) \to (1-x)^{n-1}$ for $x \to 1$.  While the analysis of
Ref.~\cite{Domokos:1971ds} assumed the completeness relation for
narrow resonances, the authors reinstated the finite width by
replacing the Dirac-$\delta$ of Eq.~(\ref{eq:wmunu-res}) into a
Breit-Wigner form and by making use of the relation $M_{J\nu n} \Gamma_{J\nu n} =
\gamma (M_{J\nu n}^2-M_{0,J}^2)$ with $\gamma=0.13$ empirically noted by Suranyi
in 1967~\footnote{We use it as quoted in Ref.~\cite{Domokos:1971ds} as
  there seems to be no reference to Suranyi's work.}  (see below for
an upgrade) accounting for the existing SLAC data at the time.  Based
on this duality picture best empirical fits to precision inclusive
electron-nucleon cross-sections data in the resonance region have been
presented (see e.g. Ref.~\cite{Christy:2007ve,Bosted:2007xd} for a
more recent analysis).

The previous discussion was restricted to vector currents.  On more
general grounds this argument holds also for any composite bilinear
current $J= \bar q {\cal O} q $ with ${\cal O}$ a Dirac spinor
operator with any quantum number connecting the nucleon with a baryon
resonance. Thus, we expect any excited baryon to asymptotically follow
the equal spacing mass squared formula, Eq.~(\ref{eq:qhd}).

\section{Quark-Diquark Models}
\label{sec:diquarks}

In the conventional quark model, baryons are made of three quarks.
However, there has traditionally been some mounting evidence that the
baryonic spectrum can be understood in terms of quark-diquark degrees
of freedom (see e.g. Ref.~\cite{Anselmino:1992vg} for a review but
also Ref.~\cite{Klempt:2017lwq} for evidence against it).  Within the
non-relativistic quark model, diquark clustering has been
investigated~\cite{Fleck:1988vm}. There exist analyses at the
Non-relativistic~\cite{Santopinto:2004hw} and
Relativistic~\cite{Ferretti:2011zz,Gutierrez:2014qpa} levels where
scalar and axial-vector diquarks have a mass of about 600 MeV and 400
MeV respectively (the diquark mass difference seems quite model
independent and about 200 MeV).  While diquarks do not resolve the
missing resonance problem, they ameliorate it since many states
predicted by the quark model do not
appear~\cite{Capstick:1986bm}. Actually, the relativistic diquark
model~\cite{Ferretti:2011zz,Gutierrez:2014qpa} does not predict
missing states below 2 GeV, whereas Isgur and Capstick have
5~\cite{Capstick:1986bm}.  On the lattice, some evidence on diquarks
correlations in the nucleon~\cite{Alexandrou:2006cq} and the dominance
of the scalar diquark channel~\cite{DeGrand:2007vu} have been
reported. More recently, within the framework of Dyson-Schwinger and
Faddeev equations, the diquark-approximation has been found to work
well~\cite{Eichmann:2016hgl}.  Radial Regge behavior in the
relativistic quark-diquark picture has been found from a numerical
analysis of the spectrum~\cite{Ebert:2011kk}.

The scaling requirement at the hadronic level implies an equidistant
mass squared spectrum for the intermediate baryonic
states/resonances. We will first argue that this condition may be
interpreted as a quark-diquark (qD) spectrum and adopt the same
argument as in the $\bar q q$ case based on the Salpeter
equation~\cite{Arriola:2006sv}. The scaling of Dirac fermions with a
heavy elementary diquark with a confining potential has also been
studied exactly and in the WKB
approximation~\cite{Jeschonnek:2005nk,VanOrden:2005zw}. If, for
simplicity, we assume scalar particles and a confining $qD$ potential,
massive diquark with mass $m_D$ the Hamiltonian in the CM frame reads
\begin{eqnarray}
M = \sqrt{p^2 + m_D^2} + p + \sigma_{qD} \, r \, , 
\end{eqnarray} 
where $\sigma_{qD}$ is a qD ``string tension''. This dynamical setup
is common to qD models (see e.g.
\cite{Santopinto:2004hw,Ferretti:2011zz,Gutierrez:2014qpa,Ebert:2011kk,DeSanctis:2014ria})
where, in addition, Coulomb, spin-spin and spin-orbital splitting
terms are added. Besides some phenomenological success in predicting
the baryonic spectrum and the familiarity with the $\bar q q$ case, we
do not know of any theoretical justification of the $qD$ linear
potential term. Good spectra are obtained for $(u,s,d)$ with $\sigma_{qD}=
2.15 {\rm fm}^{-2}$~\cite{Ferretti:2011zz}, for $(u,d)$ $\sigma_{qD}= 1.57 {\rm
  fm}^{-2}$\cite{DeSanctis:2014ria} and for $(c,b)$ with $\sigma_{qD}=
4.5 {\rm fm}^{-2}$~\cite{Ebert:2011kk}.

For excited states we can work at the semi-classical level where just
the long-distance and high-momentum configurations dominate. Indeed,
the Bohr-Sommerfeld quantization condition for $L=0$ yields (see e.g.
Ref.~\cite{Arriola:2006sv})
\begin{eqnarray}
 \oint  p_r  dr = 2 \pi (n + \alpha)   \to \frac{d n}{d M^2}= \frac1{4 \pi \sigma_{qD}}\left( 1- \frac{m_D^2}{M^2} \right) 
\end{eqnarray} 
with $\alpha $ a constant of order unity~\footnote{$\alpha$ is the
  Maslov index which depends on the smoothness of boundary conditions
  at the classical turning points (see e.g. \cite{PhysRevA.42.1907});
  the precise value is irrelevant for us here as we
  are only interested in the level density.}. For $M \gg m_D$ implies
a Radial Regge spectrum for large $n$ exactly as Eq.~(\ref{eq:qhd})
with
\begin{eqnarray}
\mu_{qD}^2 = 4\pi
\sigma_{qD} \, . 
\label{eq:sig-mu}
\end{eqnarray}
Thus, quark-diquark dynamics with a linearly confining interaction
generates excited Baryon states whose masses are asymptotically
consistent with Quark-Hadron duality in DIS.  The corresponding slopes
found in qD models are flavour dependent, $\mu_{qD}^2= 0.76 \, {\rm
  GeV}^2 $ for $(u,d)$~\cite{DeSanctis:2014ria}, $\mu_{qD}^2= 1.05 \,
{\rm GeV}^2$ for $(u,d,s)$~\cite{Ferretti:2011zz} and $\mu_{qD}^2=
2.20 \, {\rm GeV}^2$ for $(c,b)$~\cite{Ebert:2011kk}.

The role played by diquarks in high-energy processes has been
repeatedly described in the past~\cite{Anselmino:1992vg} but the
connection between this linearly confining qD interaction and DIS has
been overlooked. In our view, this connection also provides an {\it a
  posteriori} reason-of-being for recent quark-diquark models so that
their phenomenological success appears more
natural~\cite{Santopinto:2004hw, Ferretti:2011zz,Gutierrez:2014qpa}.
Nonetheless, while the absence of missing states below 2 GeV in
(relativistic) quark-diquark
models~\cite{Ferretti:2011zz,Gutierrez:2014qpa} speaks in favor of the
diquarks as effective degrees of freedom, it is unclear how the
agreement should be quantified taking into account that they are
resonances.

\section{Phenomenological Regge analysis}
\label{sec:regge}

Based on the previous discussion, for fixed $J^{PC}$ baryon quantum
numbers we propose the simple fitting formula
\begin{eqnarray}
M^2_{J\nu n} = \mu_{J\nu}^2 n + M_{0,J\nu}^2 \, . 
\label{eq:regge}
\end{eqnarray} 
In order to verify the accuracy of the radial Regge formula,
Eq.~(\ref{eq:regge}), based on the PDG listings~\cite{Olive:2016xmw}
we should provide an educated guess for the error, taking into account
that almost all states are in fact resonances having a mass $M$ and a
width $\Gamma$. Contrary to what it is often stated, the hadron
resonances listed in the PDG are, in average, narrow since the
width/mass ratio becomes $\Gamma/M=0.12(8)$ {\it both} for mesons and
baryons~\cite{RuizArriola:2010fj,Arriola:2011en} a figure which
upgrades Suranyi's result~\cite{Domokos:1971ds} and, actually,
agrees remarkably well with a salient feature of the large-$N_c$
limit of QCD where $\Gamma/M = {\cal O}
(N_c^{-1})$~\cite{Witten:1979kh} which for $N_c=3$ would numerically
give a natural value $\Gamma/M \sim 0.33$.

In order to carry out our analysis based on the 2016 PDG
compilation~\cite{Olive:2016xmw} we invoke the half-width rule (HWR)
motivated previously~\cite{Arriola:2012vk,Masjuan:2012gc}.  That way
we side-step possible channel-dependent and model-dependent
extractions of the resonance parameters assuming process-dependent
backgrounds and take into account their natural width.  Moreover, we
are then able to undertake an error analysis. The method complies to
the fact that the pole position of the resonance is typically shifted
from channel-dependent extractions by an amount $\sim \pm \Gamma
/2$. Specifically, to incorporate the half-width rule in practice, we
take along with
Refs.~\cite{RuizArriola:2010fj,Arriola:2011en,Masjuan:2012gc,Masjuan:2013xta}
the figure of merit
\begin{eqnarray}
\chi^2 = \sum_{J\nu n} \left(\frac{M_{J\nu n}^2-(M_{J\nu n}^{\rm exp})^2}{\Gamma_{J\nu n}^{\rm exp} M_{J\nu n}^{\rm exp}} \right)^2 \, ,
\label{eq:chi2}
\end{eqnarray}
in conjunction with Eq.~(\ref{eq:regge}) which corresponds to assume
that $(M_{J\nu n}^{\rm exp})^2 = M_{J\nu n}^2 \pm M_{J\nu n}
\Gamma_{J\nu n}$. Using the Suranyi relation we would obtain $(M_{J\nu
  n}^{\rm exp})^2 = M_{J\nu n}^2 ( 1 \pm 0.12(8) )$. The HWR has
already been applied to quark models fitting the charmonium spectrum
(see e.g. Appendix D of Ref.~\cite{Segovia:2011tb}), and fitting to
the unflavored light and heavy vector
mesons~\cite{Masjuan:2014sua}. In Ref. \cite{Gutierrez:2014qpa} it is
proposed to use instead {\it both} the natural width $\Gamma_{n, {\rm
    exp}}$ and the position of the resonance $\Delta M_{n, {\rm exp}}
$ in quadrature. In our case this means an inverse weight in
Eq.~(\ref{eq:chi2}) given by the replacement $\Gamma^2 M^2 \to
\Gamma^2 M^2 + 4 M^2 (\Delta M)^2 $ which reduces to ours for $\Delta
M \ll \Gamma/2$.

\begin{table}[t]
\caption{Left: slopes of radial Regge trajectories for N excited baryons, and (right) for $\Delta$ excited baryons. $\mu^2$ defined in Eq.~(\ref{eq:regge}).}
\begin{center}
\begin{tabular}{ccc| ccc}
$N,J^P$ & $\mu^2$ [GeV$^2$] & $\chi^2/{\rm DOF}$ & $\Delta,J^P$ & $\mu^2$ [GeV$^2$]& $\chi^2/{\rm DOF}$\\
\hline
$\frac12^+$ & 0.764(177) & 0.06  \\
$\frac32^+$ & 0.657(36)  & 0.01 & $\frac32^+$ & 1.061(203) & 0.01\\
$\frac52^+$ & 0.569(19) & 0.02& \\
$\frac12^-$ & 0.700(126) &  0.64 & $\frac12^-$ & 0.942(236) & 0.13\\
$\frac32^-$ & 0.640(69) & 0.16 & \\
$\frac52^-$ & 0.757(21) & 0.02 & \\
\hline
Weighted Av. & 0.657(13) & & & 1.011(153) & \\
\hline
Combined fit & 0.617(42) & 0.25 & & 1.004(153) & 0.09\\
\end{tabular}
\end{center}
\label{results}
\end{table}%
\begin{table}[t]
\caption{Left: slopes of radial Regge trajectories for N excited baryons, and (right) for $\Delta$ excited baryons from the first three lightest states. $\mu^2$ defined in Eq.~(\ref{eq:regge}).}
\begin{center}
\begin{tabular}{ccc| ccc}
$N,J^P$ & $\mu^2$ [GeV$^2$] &  $\chi^2/{\rm DOF}$ & $\Delta,J^P$ & $\mu^2$ [GeV$^2$] &  $\chi^2/{\rm DOF}$\\
\hline
$\frac12^+$ & 0.719(127) &0.15 & \\
$\frac32^+$ & 0.657(36)  & 0.01 &$\frac32^+$ & 1.061(203) & 0.01\\
$\frac52^+$ & 0.569(19) & 0.01&\\
$\frac12^-$ & 0.700(126) &  0.64 &$\frac12^-$ & 0.942(236) & 0.13\\
$\frac32^-$ & 0.566(36) & 0.03&\\
$\frac52^-$ & 0.757(21) & 0.02 &\\
\hline
Weighted Av. & 0.647(12) & & & 1.011(153) & \\
\hline
Combined fit & 0.594(46) & 0.20 & & 1.004(153) & 0.09
\end{tabular}
\end{center}
\label{results2}
\end{table}%
\begin{figure}[t]
\begin{center}
\includegraphics[width=0.4\textwidth]{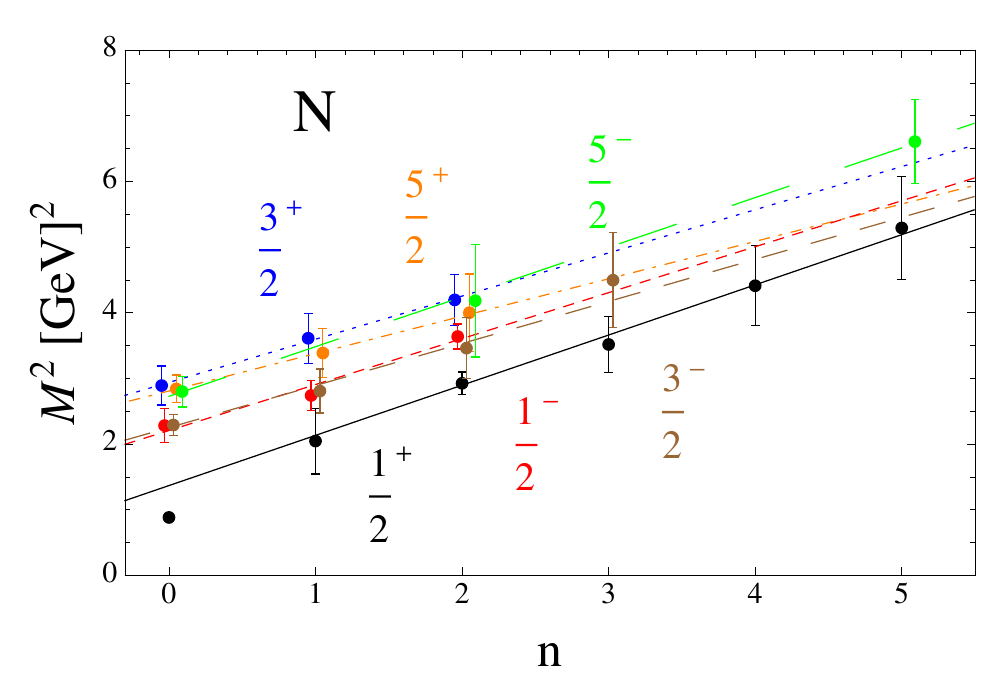}
\includegraphics[width=0.4\textwidth]{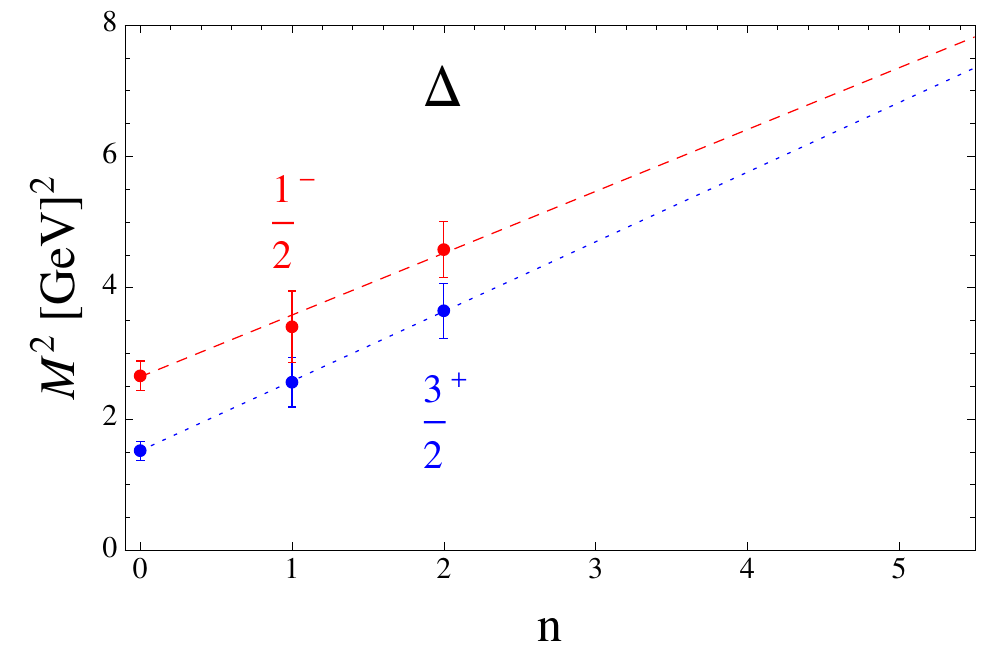}
\caption{Radial Regge trajectories for the excited N baryons (top) and
  excited $\Delta$ baryons (bottom). The $\pm$ sign indicates the
  Parity of the state: $1/2+$ is shown as solid black, $1/2-$ as red
  dashed, $3/2+$ as dotted blue, $3/2-$ as long dashed brown, $5/2+$
  as dot-dashed orange, $5/2-$ as very long dashed green.}
\label{ntraj}
\end{center}
\end{figure}

In our analysis we use $n_N=22$ and $n_\Delta=6$ states taken from the
2016 PDG compilation~\cite{Olive:2016xmw}. Our results are presented
in Figs.~\ref{ntraj} and tables \ref{results} and \ref{results2}
corresponding to fitting {\it all} states and just the {\it first
  three} states in the radial Regge trajectories respectively. As we
see, all N-states and $\Delta$-states slopes are very much alike
alike, but there is some difference between the $N$ and $\Delta$
slopes when fitted separately, $\mu_N^2 \sim 0.62 {\rm GeV}^{-2}$ and
$\mu_\Delta^2 \sim 1 {\rm GeV}^{2}$. A combined $N$ and $\Delta$ fit
with the {\it same} radial slope  using $\chi_{\rm tot}^2
\equiv \chi^2_N + \chi^2_\Delta $ produces 
\begin{eqnarray}
\mu^2 = 0.651(40) \, {\rm GeV}^2 \, , \qquad \chi_{\rm tot}^2 /{\rm DOF} = 0.49 \, . 
\end{eqnarray}
The number comes so much closer to the $N$-case since we have about
four times more $N$-states than $\Delta$-states. If we weight as to
give the same relative importance for both $N$ and $\Delta$-states as
follows
\begin{eqnarray}
  \bar \chi^2 = (n_N+n_\Delta) \left[ \frac1{n_N} \chi_N^2 + \frac1{n_\Delta} \chi_\Delta^2 \right] \, ,
  \label{eq:chi2-weight}
\end{eqnarray}
we get
\begin{eqnarray}
  \bar \mu^2 =   0.750 (32) \, {\rm GeV}^2 \, , \qquad \bar \chi^2 /{\rm
  DOF} = 1.28 \, .
\end{eqnarray}
%Using $\mu^2 = 2 \pi \sigma_{qD}$ the N case yields $\sigma_{qD}= 2.54
%{\rm fm}^{-2}$ whereas in the $\Delta$ case one gets $\sigma_{qD}=
%4.12 {\rm fm}^{-2} $. The combined $N-\Delta $ Regge fit {\it without}
%weight gives $\sigma_{qD}= 2.67(16) {\rm fm}^{-2}$ whereas the
%weighted fit using Eq.~(\ref{eq:chi2-weight}) yields $\bar
%\sigma_{qD}= 2.56(16) {\rm fm}^{-2}$.
These values should be compared with the $\mu_{qD}^2= 0.76 \, {\rm GeV}^2
$ for $(u,d)$ flavours~\cite{DeSanctis:2014ria}, whereas $\mu_{qD}^2=
1.05 \, {\rm GeV}^2$ for $(u,d,s)$~\cite{Ferretti:2011zz} and
$\mu_{qD}^2= 2.20 \, {\rm GeV}^2$ for (c,b)~\cite{Ebert:2011kk}.
Unfortunately none of these works provides uncertainties nor a value
of the $\chi^2$, but the agreement is nonetheless encouraging as we
only have (u,d) quarks.  A hybrid quark-diquark baryon model provides
{\it directly} a mass formula with splitting terms including or not
both qD as well as qqq terms~\cite{Galata:2012xt} finding
$\mu^2=1.24(5) , 1.48(5)\, {\rm GeV}^{2}$ and $\mu^2=1.50(13) \, {\rm
  GeV}^{2}$ in different scenarios.

\begin{figure}[t]
\begin{center}
\includegraphics[width=0.4\textwidth]{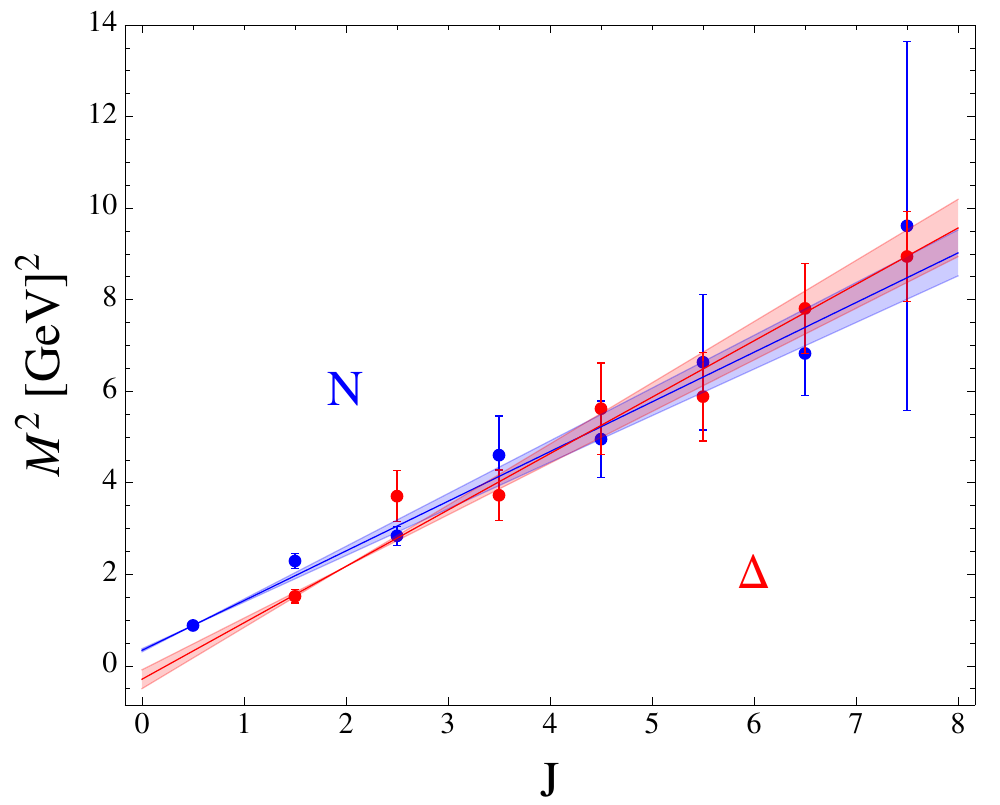}
\caption{Angular-momentum Regge trajectories. Blue trajectories for
  excited N baryons. Red trajectories for excited $\Delta$
  baryons. For N baryons, $S=1/2$, and then $J=L+1/2$. For $\Delta$
  baryons, $S=3/2$, and then $J=L+3/2$.}
\label{Jtraj}
\end{center}
\end{figure}

Motivated by quark-hadron duality, Eq.~(\ref{eq:sumint}), assuming
instead $G_{\nu n}$, when the role of $n$ and $J$ in the double
summation are interchanged we should get $ d M_{J\nu n}^2/dJ=
\beta_{\nu n}$. This suggests the angular-momentum
Regge trajectory 
\begin{equation} \label{JT}
M_{J\nu n}^2 = a_{n\nu} + \beta_{n\nu}^2 J = a_{n\nu} + \beta_{n\nu}^2 (L+S) \, , 
\end{equation}
with $S=1/2$ for excited N and $S=3/2$ for excited $\Delta$.  Results
for the fit are given in Table~\ref{results3} and are plotted in
Fig.~\ref{Jtraj}. Note that if $G_{J\nu} = G_{\nu n} = G_{\nu}$ we should
have from Eq.~(\ref{eq:sumint}) that $\mu_{J\nu}^2 = \beta_{\nu n}^2$,
which does not hold since we have $\mu^2 =0.750(30) \, {\rm GeV}^2 $
and $\beta^2=1.128(56) \, {\rm GeV}^2$.

\begin{table}[t]
\caption{Angular-momentum Regge trajectories for N  and $\Delta$ excited baryons. $\beta^2$ defined in Eq.~(\ref{JT}).}
\begin{center}
\begin{tabular}{ccc}
 & $\beta^2$ [GeV$^2$]   &  $\chi^2/{\rm DOF}$\\
\hline
N & 1.085(67) & 0.97\\
$\Delta$ & 1.232(104) & 0.72\\
\hline
Weighted Av. & 1.128(56)
\end{tabular}
\end{center}
\label{results3}
\end{table}%

\section{Comparison with other studies}
\label{sec:comp}

%\subsection{Klempt and Forkel}

The fitting equations discussed above should be compared with other
mass formulas to the PDG listings. Forkel and Klempt based on the
holographic approach~\cite{Forkel:2008un} take the mass formula,
\begin{equation} \label{dk}
M_{n,L}^2 = 4\lambda^2( n + L+3/2) - 2 (6 \lambda^2-M_N^2)\kappa\, ,
\end{equation}
\noindent
with $M_N=0.940$ GeV and $\kappa$ a good diquark fraction for each
baryon.  (In particular, $\kappa = 0$ for all $\Delta$ and spin$-3/2$
nucleon resonances, $\kappa=1/2$ for nucleons in the ground state and
$\kappa=1/4$ for the spin$-1/2$ negative-parity nucleon excitations.)
From our previous results of the combined $N$ and $\Delta$ fit on
Eq.~(\ref{eq:regge}), $\lambda =0.40(5)$GeV. However, Forkel and
Klempt, used $\lambda = 0.52$ GeV for all of them. The reason for such
a difference is basically our more restricted set of states: we
considered trajectories with at least three states which immediately
rejects those with $J>5/2$. If they were included a larger slope would
be found as it can be inferred from the result in
table~\ref{results3}. Beyond that, some states have changed since 2009
thanks to improvements in spectroscopy experiments, including those
that simply disappeared since then~\cite{Crede:2013sze}.

Still within the same classification scheme~\cite{Forkel:2008un}, one
can also consider a non-universal radial and angular-momentum
trajectories and fit the following function:
\begin{equation} \label{dk}
M_{n,L}^2 = 4\lambda_n ^2 (n+3/2)  + 4\lambda_L^2 L - 2 (6 \lambda^2-M_N^2)\kappa \, ,
\end{equation}
to get $\lambda_n =0.507(10)$ and $ \lambda_L = 0.514(11)$GeV, well compatible within errors with $\chi^2/{\rm DOF} = 0.26$. Fitting the nucleon sector exclusively, we find $\lambda_n =0.520(14)$, and $ \lambda_L = 0.512(14)$GeV, with $\chi^2/{\rm DOF} = 0.23$.

Instead of fitting the spectrum using their ansatz for the mass
square, we can use a generic $M^2_{n,L}=a +b L +c n$. We will obtain
$a=1.45(9)$GeV$^2$, $b=1.00(6)$GeV$^2$, and $c=0.83(9)$GeV$^2$, which
shows certain tension, with $\chi^2/{\rm DOF} = 0.82$.

%\subsection{Santopinto}

A similar exercise can be performed with the spectrum proposed by
Ferretti, Vassallo, and Santopinto based on solving a quark-diquark
model with a linearly confining term~\cite{Ferretti:2011zz} for the
$\Delta$ baryons. Again, a fit to $M^2_{n,L}=a +b L +c n$ yields
$a=1.53(12)$GeV$^2$, $b=1.16(34)$GeV$^2$, and $c=1.11(13)$GeV$^2$,
which shows nice agreement between slopes, with $\chi^2/{\rm DOF} =
1.36$. This is in harmony with our semi-classical argument in
Section~\ref{sec:diquarks}.

\section{Conclusions}
\label{sec:concl}

We summarize our results. Quark-Hadron duality for bilinear quark
currents connecting the nucleon with Baryon resonances provides in the
Bjorken limit a restriction on the mass spectrum of excited baryonic
states and, more specifically, on the asymptotic level density . While
the study of the Baryonic spectrum is generally and notoriously more
difficult than the Mesonic spectrum some features become remarkably
similar on the light of quark-hadron duality. The uniform mass squared
spectrum distribution befits a radial Regge-like spectrum in both
cases, and points to an effective two-body $ q \bar q$ and $q D $
dynamics for mesons and baryons respectively with a long distance
linearly growing potential. Besides, the phenomenological description
of the PDG listings of these radial Regge trajectories is satisfactory
if the finite width of the resonances is included in the error
budget. Of course, the previous arguments invoke only consistency
conditions based on high energy completeness of the baryon resonance
spectrum. More work would be needed to provide a microscopic and
dynamical justification for a dominating quark-diquark picture for
excited baryons. Conversely, we also expect some quantitative guiding
information on the slope of the radial Regge trajectories from the
analysis of structure functions with different conserved currents as
they correspond to fluctuations inside the nucleon, probing the
relative three quarks vs quark-diquark content in the excited baryon
spectrum.

\begin{acknowledgements}
PM is supported by CICYTFEDER-FPA2014-55613-P, 2014-SGR-1450 and the
CERCA Program/Generalitat de Catalunya and ERA by Spanish Mineco Grant
FIS2014-59386-P, and by Junta de Andaluc\'{\i}a grant FQM225-05.
\end{acknowledgements}

%\bibliography{../Proceedings/diquarks}
%\bibliography{diquarks}


\end{document}